\documentclass[apj,iop]{emulateapj}

\usepackage{amsmath}
\usepackage{tabularx}
\usepackage{natbib}
\usepackage{float}
\usepackage[colorlinks=true,linkcolor=blue,citecolor=blue]{hyperref}%
\usepackage[all]{hypcap}
\usepackage{lineno}
\usepackage{footnote}
\usepackage{academicons}
\usepackage[LGRgreek]{mathastext}
\usepackage[usenames,dvipsnames]{color}
\usepackage[normalem]{ulem}

\newcommand{\tint}{$T_{\rm int}$} 

\newcommand{\co}{CO}
\newcommand{\meth}{CH$_4$}
\newcommand{\amon}{NH$_3$}
\newcommand{\cotwo}{CO$_2$} 
\newcommand{\ntwo}{N$_2$} 
\newcommand{\water}{H$_2$O}
\newcommand{\phos}{PH$_3$}

\newcommand{\tp}{\emph{T--P}}
\newcommand{\rfac}{$r_{\rm fac}$}

\newcommand{\RNum}[1]{\uppercase\expandafter{\romannumeral #1\relax}}

\newcommand{\orcid}[1]{\href{https://orcid.org/#1}{\includegraphics[width=10pt]{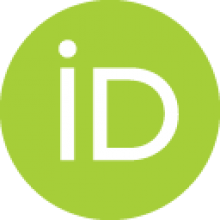}}}

\usepackage{lineno}
\shorttitle{The Atmosphere of HD 149026b from Radiative Equilibrium Models}
\shortauthors{Gagnebin et al.}

\begin{document}

\title{The atmosphere of HD 149026b: Low metal-enrichment and weak energy transport }

\email{samukher@ucsc.edu}

\author{{Anna Gagnebin$^{1}$ \orcid{0009-0003-2576-9422}},Sagnick Mukherjee$^{2}$ \orcid{0000-0003-1622-1302}, Jonathan J. Fortney$^{2}$ \orcid{0000-0002-9843-4354}, Natasha E. Batalha$^{3}$ \orcid{0000-0003-1240-6844}}
\affiliation{{$^1$}Department of Physics \& Astronomy, California State University, Sacramento, CA 95819, USA \\
{$^2$}Department of Astronomy and Astrophysics, University of California, Santa Cruz, CA 95064, USA \\
{$^3$} NASA Ames Research Center, MS 245-3, Moffett Field, CA 94035, USA \\}

\begin{abstract}
Recent {\it JWST} eclipse spectra of the high-density hot Saturn HD 149026b between 2.35 and 5.08 $\mu$m has allowed for in-depth study of its atmosphere. To understand its atmospheric properties, we have created a grid of 1D radiative-convective-thermochemical equilibrium atmosphere models and spectra with \texttt{PICASO 3.0}.  In agreement with previous work, we find that the presence of gaseous TiO creates a thermal inversion, which is inconsistent with the data. The presence of gaseous VO, however, which condenses at temperatures 200 K cooler, does not cause such inversions but alters the temperature--pressure profile of the atmosphere. We estimate an atmospheric metallicity of $14^{+12}_{-8}\times$ solar without VO and $20^{+11}_{-8}\times$ solar with VO, a factor of $\sim 10$ times 
smaller than previous work from \citet{bean23}, who relied on atmosphere retrievals. We attribute this significant difference in metallicity to a larger temperature gradient at low pressures in radiative equilibrium models.  Such models with lower metallicities readily fit the strong CO$_2$ feature at 4.3 $\mu$m.  Our lower estimated metallicity makes HD 149026b more consistent with the mass-metallicity relationship for other giant planets.  We find a C/O ratio of $0.67^{+0.06}_{-0.27}$ with and without VO. The best-fit heat redistribution factor without VO is $1.17$, a very high value suggesting very little dayside energy transport and no energy transport to the night side. The heat redistribution factor shrinks to a more plausible value of $0.91^{+0.05}_{-0.05}$, with VO, which we regard as circumstantial evidence for the molecule in the atmosphere of HD 149026b. 
\end{abstract}

\section{Introduction}\label{sec:intro}
\textit{JWST} has significantly advanced our ability to understand exoplanetary atmospheres in unprecedented detail \citep[e.g.,][]{ers_g395h,ers_nircam,ers_prism,ers_niriss,bean23,moran23,kempton23}. \textit{JWST}'s high sensitivity to infrared radiation and its coverage of infrared wavelengths from 0.6 to 29 $\mu$m makes it a fantastic observatory to study the composition and structure of exoplanetary atmospheres, as many atmospheric gases like {\water}, {\cotwo}, or {\meth} interact most strongly with infrared radiation, and it is at these wavelengths where planets emit most of their thermal flux. 


Understanding the atmospheres of highly irradiated gas giant planets is particularly important because there exists no solar system analogue of such planets. The atmospheric chemistry and temperature structure of these hot gas giants is very different than the solar system gas giants owing to their much higher equilibrium temperatures \citep{seager00,sudarsky00,fortney2005comp}. Moreover, the chemical composition of their atmospheres can help us
to determine their fundamental atmospheric properties such as metallicity and the carbon-to-oxygen (C/O) ratio \citep[e.g.,][]{madhu12,fortney13,ers_g395h,ers_nircam,ers_niriss,ers_prism}. These fundamental atmospheric parameters can contain important clues about the formation and evolutionary history of planets and can also help us in solving the mystery of how such giant planets are found in such proximity to their host stars \citep[e.g.,][]{oberg11,fortney13,mordasini16,molliere22}. HD 149026b is one such hot Saturn planet which was recently studied with {\it JWST} using eclipse spectroscopy by \citet{bean23}.

The planet HD 149026b was detected by \citet{sato2005} with {\it Subaru} and {\it Keck} radial velocity observations of the host star. Photometric follow-up of the system by \citet{sato2005} revealed transits of the planet and enabled precise measurements of its radius. HD 149026b was found to be unusually dense, which further suggested that it might have a high bulk metallicity \citep{sato2005,fortney06}. Since then, many follow-up studies have resulted in constraints on properties for both the planet and its host star \citep[e.g.,][]{torres08,Carter_2009,southworth10,albrecht12,knutson14,bonomo17,zhang18_hd149206,ment18,stassun17}. For this work we use the stellar and planetary properties of HD 149026b reported in \citet{bean23}. HD 149026b has a radius $R_{p} = 0.723 \pm 0.029$ $R_{Jup}$, a mass $M_p = 0.358 \pm 0.018$ $M_{Jup}$, an equilibrium temperature $T_{eq} = 1694$ K (assuming zero Bond albedo), orbital period of $P = 2.87588874$ days, and orbital eccentricity of $e = 0.0$ \citep{bean23}. The host star is G0 IV spectral type \citep{sato2005} with a mass of $M_{s} = 1.28 \pm 0.08$ $M_{\odot}$, a radius of $R_{s} = 1.454 \pm 0.048$ $R_{\odot}$, an effective temperature of $T_{eff} = 6085\pm100$ K, and a stellar metallicity significantly enriched compared to the Sun,  with $[Fe/H] = 0.25\pm0.10$ \citep{bean23}.

In the years since its discovery, HD 149026b has been a target of many modeling and atmospheric characterization studies.  \citet{fortney06} identified the possibility that HD 149026b belongs to a parameter space where it might develop stratospheric temperature inversions due to gaseous TiO and VO in its atmosphere, building off of work from \citet{hubeny03}. The secondary eclipse of HD 149026b was observed with {\it Spitzer} at 8 $\mu$m by \citet{harrington07}. This data was later re-analyzed by \citet{knutson09}, which found a day-side brightness temperature of 1440 $\pm$ 150 K for the planet. Significant phase-curve variation for the planet was also seen at 8 $\mu$m with {\it Spitzer} by \citet{knutson09}, indicating the presence of a strong day-night temperature contrast. 

\citet{stevenson12} reported measurements of the eclipse depths of HD 149026b at multiple {\it Spitzer} channels covering 3.5 to 16 $\mu$m. These measurements ruled out the presence of a thermal inversion in the planet's atmosphere and suggested that the planet's atmosphere to be significantly metal-rich ($\sim$ 30$\times$ solar metallicity). \citet{line14} performed a retrieval analysis on the measurements from \citet{stevenson12} and found a C/O ratio only modestly constrained, from 0.45-1.0. The lack of an inversion was further substantiated by \citet{ishizuka21} when no TiO was detected in the planet's atmosphere but neutral Ti atoms were observed with high-resolution spectroscopy. \citet{zhang18_hd149206} measured the phase-curves of HD 149026b with {\it Spitzer} in the 3.6 and 4.5 $\mu$m bands and found that while 3D circulation models could explain some aspects of the phase curves, several discrepancies between phase curve data and 3D GCMs remain. 

Most recently, \citet{bean23} used the \textit{JWST} Near Infrared Camera (NIRCam) to obtain an eclipse/emission spectrum of HD 149026b between 2.3-5.1 $\mu$m. \citet{bean23} performed a chemically consistent Bayesian retrieval analysis in order to determine the characteristics of the atmosphere. Using this method, \citet{bean23} found that HD 149026b has an atmospheric metallicity between 59-276 $\times$ solar and a C/O ratio of $0.84 \pm 0.03$. Along with gaseous abundances, the atmospheric temperature-pressure ({\tp}) profile itself is a very important parameter in understanding the depth-dependent atmospheric energy balance, and in  interpreting the emission spectra of planets.  In conjunction with the atmospheric abundances, \citet{bean23} used the {\tp} profile parametrization from \citet{line13} to retrieve the {\tp} profile of the planet. The metallicity estimates from \citet{bean23} indicated that HD 149026b has a much higher atmospheric metallicity than estimated from the mass-metallicity relationship constrained by solar system gas giant planets and hot Jupiters.

While Bayesian atmospheric retrievals are extremely useful in putting constraints on gaseous abundances and {\tp} profiles of planets, they can also be very sensitive to the presence of minor systematics within the data and may also retrieve solutions that are physically or chemically inconsistent. 1D radiative--convective equilibrium models, on the other hand, are physically consistent based on a set of assumptions, e.g., chemical equilibrium/disequilibrium (see \citet{marleyrobinson15}), but given constraints mentioned will typically not achieve fits that are statistically as good as those found from retrievals.

It has become clear that no one modeling technique is enough to understand the true nature of exoplanetary atmospheres, and results from each technique must always be put into context with another. For example, retrieved chemical abundances must be analyzed in the context of abundances expected from 1D radiative-convective models to check whether the abundance of a gas retrieved in a planet is consistent with the predictions from 1D radiative-convective models. Similarly, {\tp} profiles from 1D radiative-convective models must be checked against retrieved {\tp} profiles to determine the source of the differences. These may be due to a missing heating/cooling mechanism or opacity source in the radiative-convective model or may be a result of a specific methodology or data systematics in the retrieval.

Over the past two decades, various aspects of 1D radiative-convective models for planets have been improved based on observations of both exoplanets and brown dwarfs. Therefore, in this work, we analyze the eclipse spectrum of HD 149026b reported by \citet{bean23} using self-consistent 1D radiative-convective-thermochemical equilibrium models and compare our results with the retrieval analysis performed by \citet{bean23}. A major goal of this work is to reassess the highly metal enriched atmosphere found with retrieval studies of HD 149026b in \citet{bean23} with 1D radiative-convective models. We also discuss the various similarities and differences between the results from the two methods and what those can tell us about HD 149026b's atmosphere.

We briefly describe the observations and data reduction procedure followed by \citet{bean23} in \S\ref{sec:observations} followed by a detailed description of our 1D radiative-convective model grid in \S\ref{sec:model}. In \S\ref{sec:gridtrievals}, we describe our model fitting technique, and in \S\ref{sec:results}, we report our key results. This is followed by a discussion of our results in \S\ref{sec:discussion} and the main conclusions of this work in \S\ref{sec:conclusion}.

\section{{\it JWST} NIRCAM Observations}\label{sec:observations}
\citet{bean23} used the \textit{JWST} Near-Infrared Camera (NIRCam) to determine the planet-to-star flux ratio at secondary eclipse for HD 149026b. These observations were made over 8.27 hours on July 15 and August 4, 2022 during the secondary eclipse (or occultation), as the planet passed behind the host star. On the first observation, they used the NIRCam F322W2 filter and looked between 2.349 and 4.055 $\mu$m. On the second observation, they used the NIRCam F444W filter between 3.778 and 5.082 $\mu$m. This results in a continuous thermal emission spectrum between 2.349 and 5.082 $\mu$m.

To collect the near-infrared spectra as a function of time, both observations used the module A grism R mode, whose function is described in \citet{Greene_2017}. \citet{bean23} then used the \texttt{Eureka!} pipeline to reduce the data into a simple spectrum. The details of the \texttt{Eureka!} data reduction pipeline can be found in \citet{Bell_2022}. 

\section{Atmospheric Modeling}\label{sec:model}
We use the \texttt{PICASO} 1D radiative--convective--thermochemical equilibrium (RCTE) atmospheric model to compute models of HD 149026b's atmosphere \citep{Mukherjee22,batalha19}. \texttt{PICASO} is a well-vetted Python model that has its legacy from the FORTRAN based \texttt{EGP} code which has been widely used to model atmospheres of Solar System objects \citep[e.g.,][]{mckay1989thermal,Marley96}, exoplanetary atmospheres \citep[e.g.,][]{fortney2005comp,fortney2007planetary,fortney2008unified,fortney20}, and brown dwarf atmospheres \citep[e.g.,][]{marley21,morley14water,karilidi2021,Mukherjee22a}. The \texttt{PICASO} model has been benchmarked against these applications in \citet{Mukherjee22} and has been since used to model {\it JWST} observations of exoplanets like WASP-39b \citep[e.g.,][]{ers_g395h,ers_nircam,ers_niriss,ers_prism} and several brown dwarfs \citep[e.g.,][]{Greenbaum2023,miles22,beiler23}.

We divide HD 149026b's atmosphere into 91 plane-parallel pressure layers with logarithmically spaced pressure from the deep convective atmosphere with highest pressure of 300 bars to the upper radiative atmosphere, out to $10^{-7}$ bars. We use the measured stellar and planetary parameters for HD 149026b reported in \citet{bean23} and \citet{brewer16} as input parameters for atmospheric modeling with \texttt{PICASO}. The required stellar input parameters are stellar $T_{\rm eff}$, metallicity, $log(g)$, and radius. The planetary mass, radius, star-planet separation, intrinsic temperature ($T_{\rm int}$), atmospheric metallicity, C/O ratio, and atmospheric heat recirculation factor (\rfac, described later) are also needed as input parameters to create atmosphere models for the planet.


We fix the stellar parameters, planetary mass, planetary radius, and star-planet separation to their measured values. Table \ref{tab:table1} lists the values of these parameters used in this work and their respective sources. As we aim to constrain the planetary atmospheric parameters -- $T_{\rm int}$, atmospheric [M/H], C/O, and heat recirculation factor -- we let them vary across an extensive range of values to create a large grid of  2160 forward model atmospheres. The range of these parameters are shown in Table \ref{tab:table1}.

\begin{table*}\label{tab:table1}
\begin{center}

 \begin{tabular}{|c || c | c | c ||} 
 
 \hline
 {\bf Parameter} & {\bf Used Value} & Increment & {\bf Reference} \\ [0.5ex] 
 \hline\hline

 Stellar $T_{\rm eff}$ & 6085 K & Fixed & \citet{brewer16,bean23} \\
  \hline
 Stellar $log(g)$ & 4.24 [cgs] & Fixed & \citet{brewer16,bean23} \\
  \hline
 Stellar [M/H] & 0.25 dex & Fixed & \citet{bean23} \\
  \hline
 Stellar R & 1.454 R$_{\odot}$ & Fixed & \citet{bean23} \\
 \hline
 $M_{p}$ & 0.358 $M_{J}$ & Fixed & \citet{bean23} \\
 \hline
 $R_{p}$ & 0.723 $R_{J}$ & Fixed & \citet{bean23} \\
 \hline
 a & 0.0436 AU & Fixed & \citet{bean23} \\
 \hline
 $T_{\rm int}$ & 100-300 K & 200 K & -- \\
 \hline
 [M/H] & -1.0 to +2.0 & $\sim$ 0.2/0.3 & -- \\
 \hline
 C/O & 0.11-0.916 & $\sim$ 0.25/0.5 & -- \\
 \hline
 \rfac\ & 0.5-1.3 & $\sim$ 0.1 & -- \\
 \hline
\end{tabular}
\end{center}
\caption{System and atmospheric parameters of the exoplanet HD 149026b and its host star used in this work.}
\label{table:table1}
\end{table*}

Using the stellar and planetary parameters as well as an initial guess for the atmospheric temperature-pressure (\tp) structure, \texttt{PICASO} numerically iterates the chemical composition, \tp\ profile, and locations of radiative and convective zones until radiative--convective equilibrium is achieved throughout the atmosphere. During the model iterations, the thermochemical equilibrium abundances of gases are interpolated from precalculated chemistry tables using the methods of \citet{visscher05,visscher06,visscher10}, as updated in \citet{marley21}. The correlated-k opacities of individual gases including {\co}, {\meth}, {\water}, {\amon}, {\ntwo}, {\cotwo}, HCN, H$_2$, {\phos}, C$_2$H$_2$, Na, K, and FeH are obtained from \citet{lupu_roxana_2021_7542068} and mixed ``on--the--fly" in 661 wavelength bins during each iteration using the method described in \citet{amundsen17}. 

The converged \tp profile and the atmospheric chemical abundances are used to calculate the emission spectrum of the exoplanet. The emission spectra are created at a spectral resolution of 60,000 and then binned down to the data  resolution for comparison. We use the stellar spectral models from \citet{castelli2004grid} along with \texttt{PySynPhot} \citep{pysynphot2013} to calculate planet-to-star flux ratio ($F_{\rm planet}$/$F_{\rm star}$). The \citet{castelli2004grid} stellar atmospheric models were used to maintain uniformity with the analysis done in \citet{bean23}. 
In the following subsections, we show how each of our key atmospheric parameters -- atmospheric metallicity, C/O ratio, heat re-circulation factor, and intrinsic temperature, changes the \tp profiles and the F$_{\rm planet}$/F$_{\rm star}$ spectra for HD 149026b.
\begin{figure*}
    \centering
    \includegraphics[width = \textwidth]{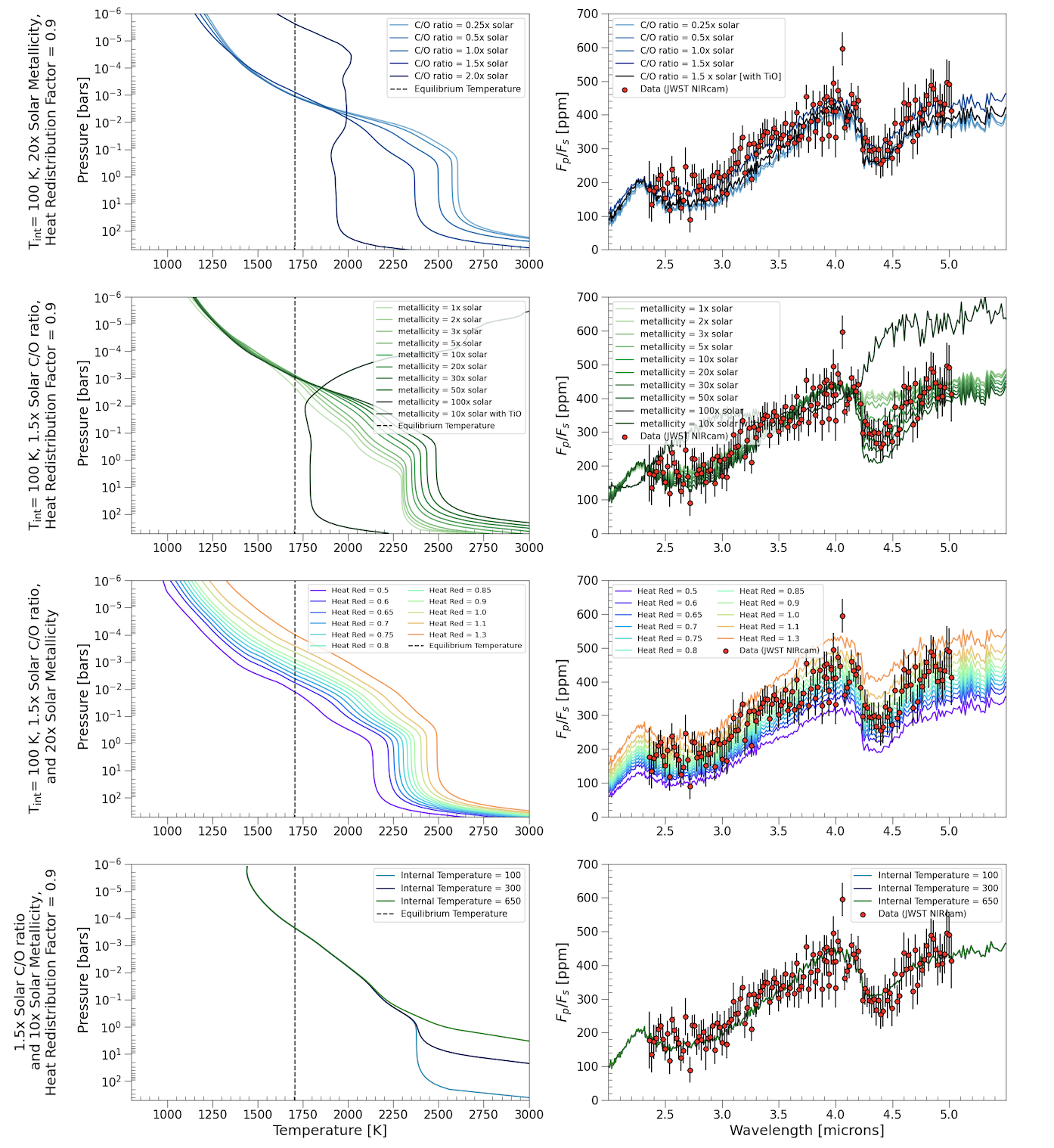}
    \centering
    \caption{Models of HD 149026b with a range of atmospheric properties. The panels on the left show the temperature-pressure (\tp) profiles, with a black dashed line showing where the temperature is equal to the equilibrium temperature of the planet. The panels on the right show the model thermal emission spectra, which are calculated by dividing the model spectrum of the planet by the spectrum of the star. The \textit{JWST} NIRCam spectral data is shown in red. The top row shows models with \tint\ 100K, 20$\times$ solar metallicity, and \rfac\ of 0.9 with varying C/O ratio. 
    The second row shows models at \tint\ 100 K, a C/O ratio of 1.5$\times$ solar, and \rfac\ of 0.9 with varying metallicities. There's an additional model in this row which includes gaseous TiO in the atmosphere as well as VO. This model shows a strong temperature inversion in the upper atmosphere as well as a clear mismatch to the data, leading to our decision to not include TiO in our grid of models. 
    The third row shows models at \tint\ of 100K, a C/O ratio of 1.5$\times$ solar, and a metallicity of 20x solar with varying \rfac\ values. 
    The final row shows models at 1.5$\times$ solar C/O ratio, 10$\times$ solar metallicity, and \rfac\ of 0.9 with varying intrinsic temperatures. 
    It is clear that the precise {\it JWST} NIRCAM observations can constrain the C/O ratio, metallicity, and the heat redistribution factor, while constraining {\tint} will perhaps not be possible, as it does not vary  the spectrum in any appreciable way, as seen in the fourth row}.
    \label{fig:models}
\end{figure*}

\subsection{Carbon to Oxygen Ratio}

The C/O ratio controls the relative abundances of C- and O- carrying gases in the atmosphere \citep{madhu12}. The abundances of a variety of gases such as {\meth}, {\co}, {\water}, and {\cotwo} are strongly influenced by the C/O ratio. A C/O ratio approaching $\sim 1$ indicates that the atmosphere starts to be dominated by C- bearing gases like {\meth} and HCN  while the abundance of O- bearing gases like {\water}, {\cotwo}, and {\co} are relatively low due to limited availability of O- atoms \citep[e.g.,][]{madhu12,molliere15,goyal18}. Inversely, a C/O ratio much less than 1 increases the relative abundances of O- bearing gases like {\water},{\cotwo}, and {\co} while {\meth} abundances are lower due to a limited supply of carbon atoms. Figure \ref{fig:models} top left and right panels show the effect of varying C/O ratio on the \tp profile (left panel) and the observable spectra (right panel) in a super-solar metallicity (20$\times$) atmospheric model for HD 149026b. The C/O ratio has been varied from sub-solar values of 0.11 to super-solar values of 0.916 here, where 0.458 has been assumed to be the solar C/O \citep{lodders09}. 
The top left panel of Figure \ref{fig:models} shows that at $T_{int} = 100$ K, an atmospheric metallicity of $20\times$ solar, and a heat redistribution factor of 0.9, the C/O ratio has a strong influence on the atmospheric \tp especially at pressures greater than $\sim$ 10$^{-2}$ bars. Figure \ref{fig:models} shows that the \tp profile gets colder at pressures greater than $\sim$ 10$^{-2}$ bars with increasing C/O ratio. As the C/O increases, the abundance of gases like {\water}, {\cotwo}, and {\co} decreases, which causes the atmosphere to be colder in the deeper parts of the atmosphere. However, a thermal inversion develops once the C/O ratio reaches a high value, here 0.916. This behavior has been explored previously in \citet{molliere15} and is due to the drastically changing atmospheric chemistry near C/O values of 0.9. Near C/O values of 0.9, the atmosphere is neither very rich in O- bearing gases like {\water} nor C- bearing gases like HCN or {\meth}. As a result, the upper atmosphere lacks these coolants but the alkali metals (Na and K) continue absorbing the host star light and strongly heats the upper atmosphere, causing a thermal inversion.  While the corresponding model spectra calculated at a range of C/O ratios agree with the observations to varying degrees, a C/O ratio of 0.916 is clearly ruled out as the nearly isothermal atmosphere leads to a featureless spectrum.

\subsection{Metallicity}
As the metallicity of an atmosphere increases, the atmospheric abundances of gases like $CO$ and $CO_2$ also increase \citep{Lodders_Fegley_2002}. Moreover, CO increasingly becomes a favored C- carrier over {\meth} with increasing metallicity under thermochemical equilibrium \citep{Lodders_Fegley_2002}. The left and the right panels in the second row of Figure \ref{fig:models} show the effect of varying metallicity on the atmospheric \tp profile and spectra, respectively. As metallicity increases, optical depths typically increase at all model levels, pushing the photospheric levels to lower pressure.  The increased thermal opacity leads to hotter atmospheres. The left panel ($T_{int} = 100$ K, $1.5\times$ solar C/O ratio, and a heat redistribution factor of 0.9), shows that this warming of the \tp profile with increasing metallicity mainly appears at pressures greater than 10$^{-3}$ bars in models of HD 149026b's atmosphere.

The right panel shows the dramatic effect of metallicity on the {\cotwo} feature between 4-4.2 $\mu$m.  This feature cannot be matched with models having metallicities lower than $\sim5\times$ solar. The {\cotwo} feature is expected to provide the strongest constraints on metallicity as it is very sensitive to this parameter.

\subsection{Heat Redistribution Factor}
The heat redistribution factor parameter (\rfac) is a value used in 1D atmosphere models to describe the redistribution of absorbed stellar energy. Due to their proximity to their host stars, transiting exoplanets are expected to be tidally locked to their star. This can create large temperature contrasts between the permanent day and night side along with the presence of strong day to night winds. Such winds can transport part of the energy from the day to the night, thus effectively somewhat cooling the day side and heating the night side.

In this work, we model the irradiated day hemisphere. The \rfac\ parameter is varied between 0.5 to 1.3. A value of 0.5 corresponds to the scenario where the flux received from the host star is re-radiated into space from both the day and night hemispheres of the planet due to complete heat redistribution \citep{hansen08}. On the other hand, a value of 1.33 corresponds to no redistribution of energy at all, where the dayside emitted flux is dominated by the hottest point at ``high noon" on the planetary dayside.  Within this framework, a value of 1.0 means that still no flux is lost to the night side, but the day side homogenizes itself \citep{hansen08}. This heat redistribution parameter is a very approximate way to capture the 3-dimensional nature of exoplanet atmospheres within 1D models. We note that the numerical range of this factor is often defined differently by different 1D atmospheric models for exoplanets \citep[e.g.,][]{hansen08,goyal18}.

The left column in the third row of Figure \ref{fig:models} shows the effect of varying \rfac\ from 0.5 to 1.3 at a $T_{int}$ of 100K, a C/O ratio of $1.5\times$ solar, and a $20\times$ solar metallicity. A higher redistribution factor causes a warmer \tp profile for the atmosphere along with a higher thermal flux, as shown in the right panel in the third row of Figure \ref{fig:models}. The comparison of these models with varying heat redistribution with the observed data makes it clear that low heat redistribution factors ($<$ 0.7) are not favored. 

\subsection{Intrinsic Temperature}
The magnitude of flux emitted from the interiors of close-in giant planets is still an open question \citep{Thorngren_2019empirical}. The intrinsic temperature ($T_{\rm int}$) quantifies the planet's interior heat flux as $F_{\rm int}= \sigma T_{\rm int}^4$. HD 149026b will have a much higher $T_{\rm eq}$ than $T_{\rm int}$ due to its proximity to its host star. As a result, its upper atmosphere \tp profile and chemistry are expected to be influenced by $T_{\rm eq}$ alone and not $T_{\rm int}$. However, $T_{\rm int}$ controls the deeper atmosphere \tp profile and the adiabat in the deep convective region of the planet \citep{fortney2007planetary,fortney20}. 

Even though we do not expect the {\it JWST} observations of HD 149026b to be very sensitive to $T_{\rm int}$, we have allowed a very coarse variation of $T_{\rm int}$ in our atmospheric model grid between 100 K and 300 K. The last row of Figure \ref{fig:models} shows the effect of varying $T_{\rm int}$ on the atmospheric \tp profile and the spectra. It is clear that $T_{\rm int}$ does not affect the model spectra shown in the right panel, but it does significantly change the \tp profile at pressures higher than $\sim$ 0.5 bars and the deep convective adiabat. We also included a model at 650 K to show that adding an additional grid point at a higher temperature would not affect our grid-retrieval.

\subsection{Models With and Without Gaseous VO}
HD 149026b has a $T_{\rm eq}$ (1694 K) such that it resides near the boundary of a proposed classification of hot Jupiter atmospheres into the ``pM" and ``pL" classes depending on whether strong optical absorbers -- TiO and VO are present (pM), or condensed out (pL), in analogy with M and L dwarfs \citep{fortney08,fortney06,hubeny03}. Including both TiO and VO in gaseous forms while calculating the \tp profiles with \texttt{PICASO} leads to very strong thermal inversions for HD 149026b's model atmospheres, as shown in Figure \ref{fig:models}. Figure \ref{fig:models} also shows that the spectrum of \citet{bean23} does not support an inversion for this planet.  This could be due to a deep atmosphere cold trap \citep{fortney08,spiegel09}, where the deep \tp\ profile crosses a condensation curve for a Ti-bearing condensate.  Moreover, ground-based high-resolution spectroscopy of HD 149026b has not detected the presence of gaseous TiO \citep{ishizuka21}.

\begin{figure}
    \centering
    \includegraphics[width = 0.5\textwidth]{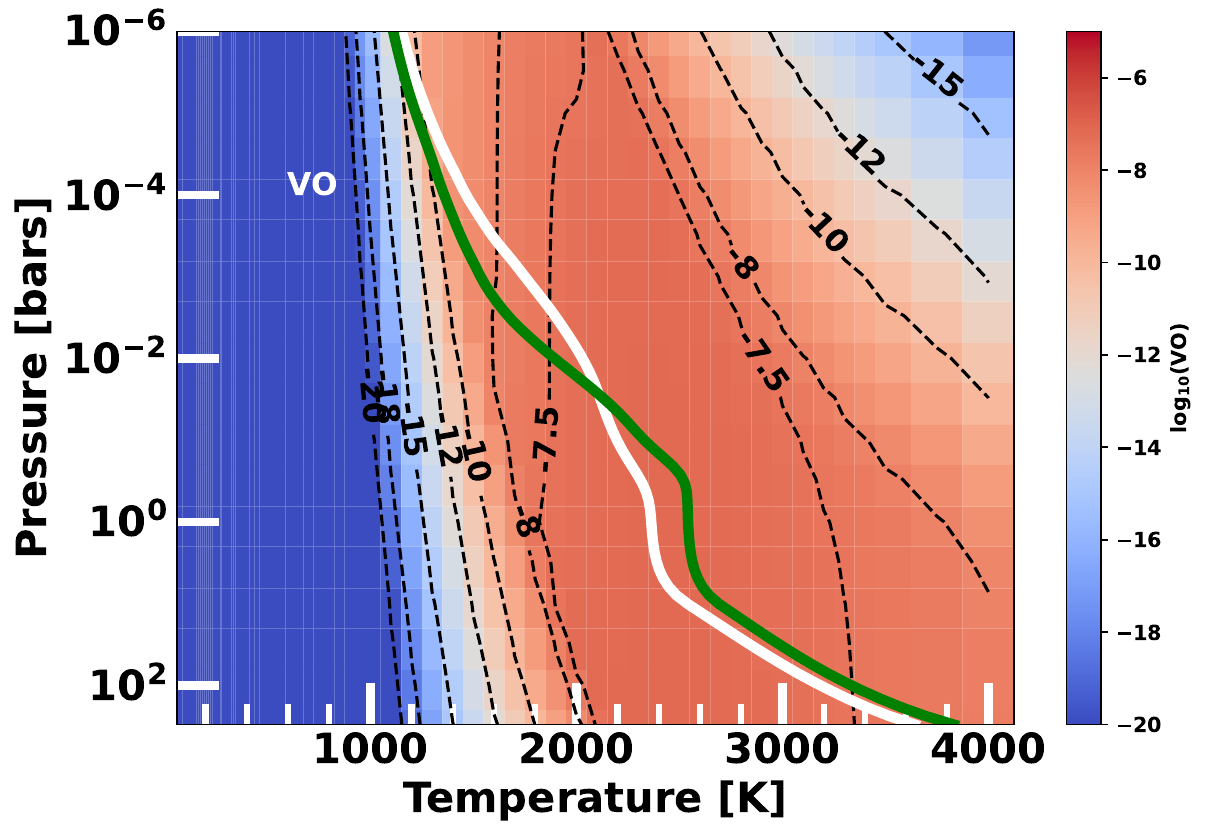}
    \centering
    \caption{Heat map showing the expected volume mixing ratio of gaseous VO as a function of pressure and temperature under thermochemical equilibrium. The map presented here is for 10$\times$solar metallicity and a C/O of 0.667. The white line shows a model \tp profile from our model grid for HD 149026b which includes gaseous VO whereas the green line represents a model \tp calculated without gaseous VO. The sharp drop in VO abundance towards lower temperatures is a result of condensation into V-bearing condensates.}
    \label{fig:VO_map_models}
\end{figure}

However, VO is a comparatively milder absorber (due to a lower abundance) at the same optical wavelengths where TiO absorbs.  As discussed in \citet{fortney08}, the condensation curve of V is $\sim~200$ K cooler than of Ti, such that VO could still be present in the atmosphere of HD 149026b in gaseous form. Figure \ref{fig:VO_map_models} shows a heat map of gaseous VO abundance as a function of temperature and pressure for 10$\times$solar metallicity and C/O of 0.667. The gas phase VO abundance in Figure \ref{fig:VO_map_models} shows a sharp drop at colder temperatures due to condensation of V. 

Therefore, to explore this behavior, we create two sets of model grids for HD 149026b -- one without gaseous VO and one with gaseous VO, neither of which included gaseous TiO. An example of each of these two cases are shown with corresponding \tp profiles overplotted on the gas phase VO abundance heat map in Figure \ref{fig:VO_map_models}. The white line shows a \tp profile for a 10$\times$solar metallicity atmosphere calculated with gaseous VO whereas the green line is a profile calculated without any gaseous VO. Figure \ref{fig:VO_map_models} makes it clear that even though TiO has not been detected in HD 149026b's atmosphere, there is a possibility of significant amount of gaseous VO remaining in its atmosphere. 

\begin{figure*}
    \centering
    \includegraphics[width = \textwidth]{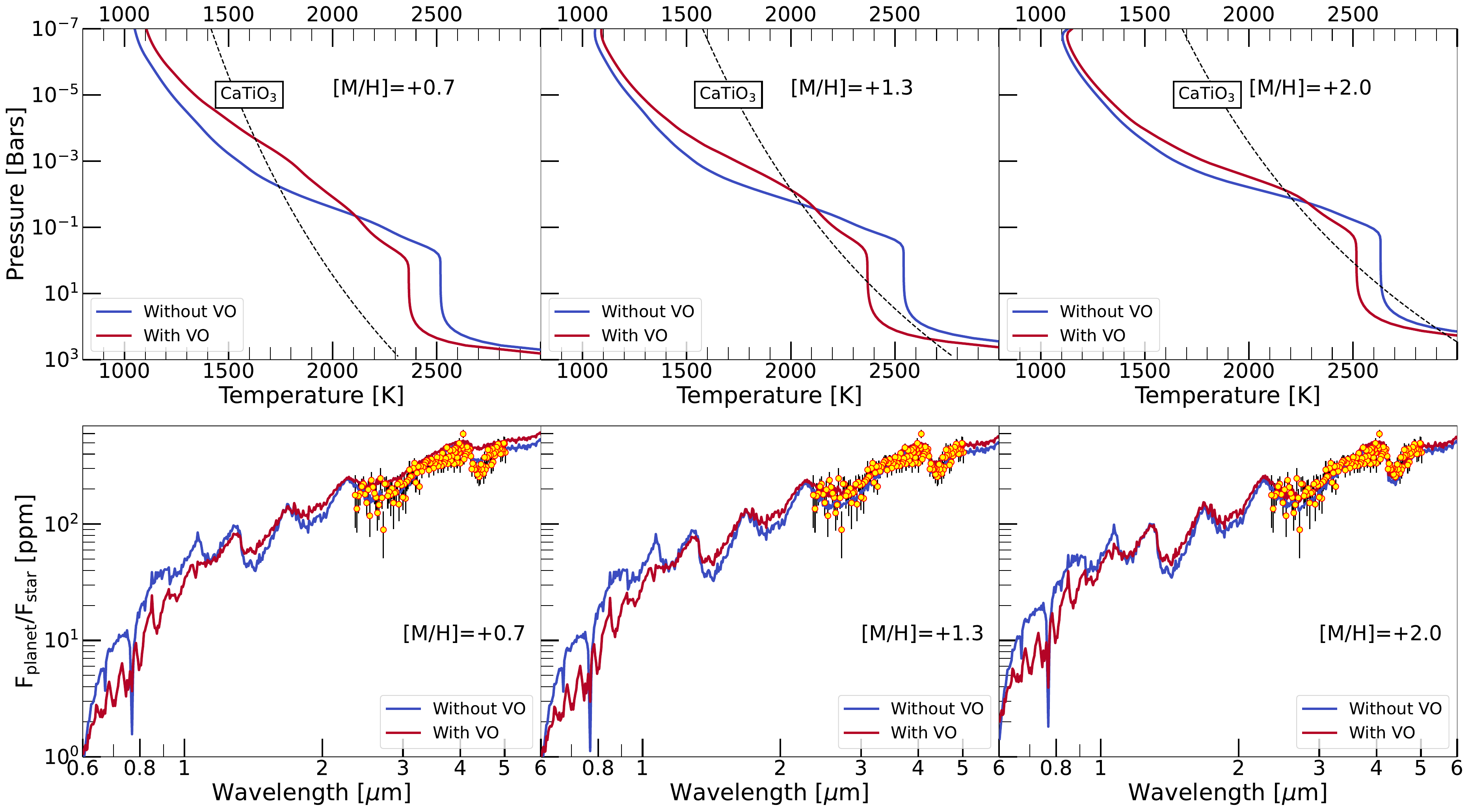}
    \centering
    \caption{The top three panels show  model temperature-pressure profiles for HD 149026b at three different metallicity values. The blue line shows what the \tp\ profile without  VO in the atmosphere and the red line shows the profile with VO. For reference, the condensation curve of the main Ti-bearing condensate (CaTiO$_3$, which removes gaseous TiO) is shown as a dashed black curve.  The bottom panels show the corresponding thermal emission spectra. The wavelength-coverage of the {\it JWST} NIRCAM data makes it difficult to differentiate between the scenarios with and without gaseous VO. However, observations in shorter wavelengths can be very useful to differentiate between these two scenarios.}
    \label{fig:w_wo_VO_models}
\end{figure*}

The top panels of Figure \ref{fig:w_wo_VO_models} show the effect of including gaseous VO in the atmosphere of HD 149026b on the atmospheric \tp profile at various metallicities. We find that VO, unlike TiO, does not create a thermal inversion in these models when added at abundances predicted from chemical equilibrium. However, it still has a significant effect on the \tp profile throughout the atmosphere. By absorbing more stellar flux in the upper atmosphere due to its large cross-section at optical wavelengths, gaseous VO warms the upper atmosphere at pressures less than $\sim$ 10 mbars, compared to a case without gaseous VO. However, this effect reverses itself at pressures greater than 10 mbars, where the atmosphere without gaseous VO is hotter than the atmosphere with gaseous VO. The presence of gaseous VO in the atmosphere can also reduce the Bond albedo of the planet by a factor of $\sim$3 from a typical value of $\sim$0.03 to $\sim$0.01 in cloud-free atmospheres.

The bottom panels of Figure \ref{fig:w_wo_VO_models} show the spectra computed from each model from the top panels compared with the {\it JWST} observations. The presence or absence of gaseous VO in the planet's atmosphere cannot be discerned with the identification of distinct VO absorption features in the wavelength range of the {\it JWST} observations. However, the emission spectra of planets are also sensitive to their \tp profiles. Therefore we include the presence/absence of gaseous VO within our analysis as it can have large impacts on the \tp profile of HD 149026b. The bottom panels of Figure \ref{fig:w_wo_VO_models} also show that observations of this planet in shorter wavelengths will be useful to distinguish between the two scenarios.  

\section{Grid Fitting}\label{sec:gridtrievals}
For a more quantitative assessment of the atmosphere of HD 149026b we use a grid-retrieval technique to fit the \citet{bean23} spectra with our forward model grids. This technique was used to model the transmission spectrum data of hot-Saturn WASP-39b by the {\it JWST} transiting planets Early Release Science program \citep{ers_prism,ers_g395h,ers_nircam,ers_niriss} as well as the thermal emission spectrum of WASP-18b (\citet{Coulombe_2023}, \cite{Brogi_2023}, \citet{Arcangeli_2018}) and HAT-P-7b (\citet{Mansfield_2021}. The \tp profile and the atmospheric chemical abundance profiles are first computed on the chosen grid points of metallicity, C/O, heat redistribution factor, and $T_{\rm int}$ with a common pressure grid that is appropriate for all the models. To sample parameter values between the grid points, we linearly interpolate the temperature for each pressure layer as a function of atmospheric pressure, metallicity, C/O, heat redistribution factor, and $T_{\rm int}$ using the \texttt{scipy.interpolate.RegularGridInterpolate} routine. Similarly, the logarithm of the chemical abundances of all the atmospheric gases within our models are also interpolated using the same routine as a function of atmospheric pressure and the other varying parameters described above.

We fit the observed data with our model grid using the Dynamic Nested Sampling tool \texttt{DYNESTY} \citep{speagle20}. For each iteration within the nested sampler, the interpolation function for the temperature at each atmospheric pressure is called with the relevant parameters drawn from the parameter space by the nested sampler. This procedure produces an interpolated \tp profile at the point of the parameter space chosen by the sampler. Similarly, the interpolated functions for H$_2$, He, {\water}, {\meth}, {\co}, {\cotwo}, {\amon}, {\ntwo}, Na, K, H$_2$S, and C$_2$H$_2$ are also called for each iteration to compute the interpolated chemical abundance profiles at the point of the parameter space drawn by the sampler.

The resulting $F_{\rm planet}$/$F_{\rm star}$ spectrum of the planet is then computed using the interpolated \tp and chemical abundance profiles. This model spectrum is binned to the wavelength points of the observed spectrum and is then used to compute the log-likelihood function for the sampled point in the parameter space. The nested sampler continues sampling the parameter space with these iterations and tries to reduce the log-likelihood function until a predetermined convergence criterion is met. For estimating our four parameters of interest, we use 600 live points for our nested sampler. We use uniform priors on $T_{\rm int}$, C/O, and heat redistribution function while uniform priors are used for the logarithm of the atmospheric metallicity.

\section{Results}\label{sec:results}

\subsection{Best Fitting Emission Spectra}
\begin{figure*}
    \centering
    \includegraphics[width = \textwidth]{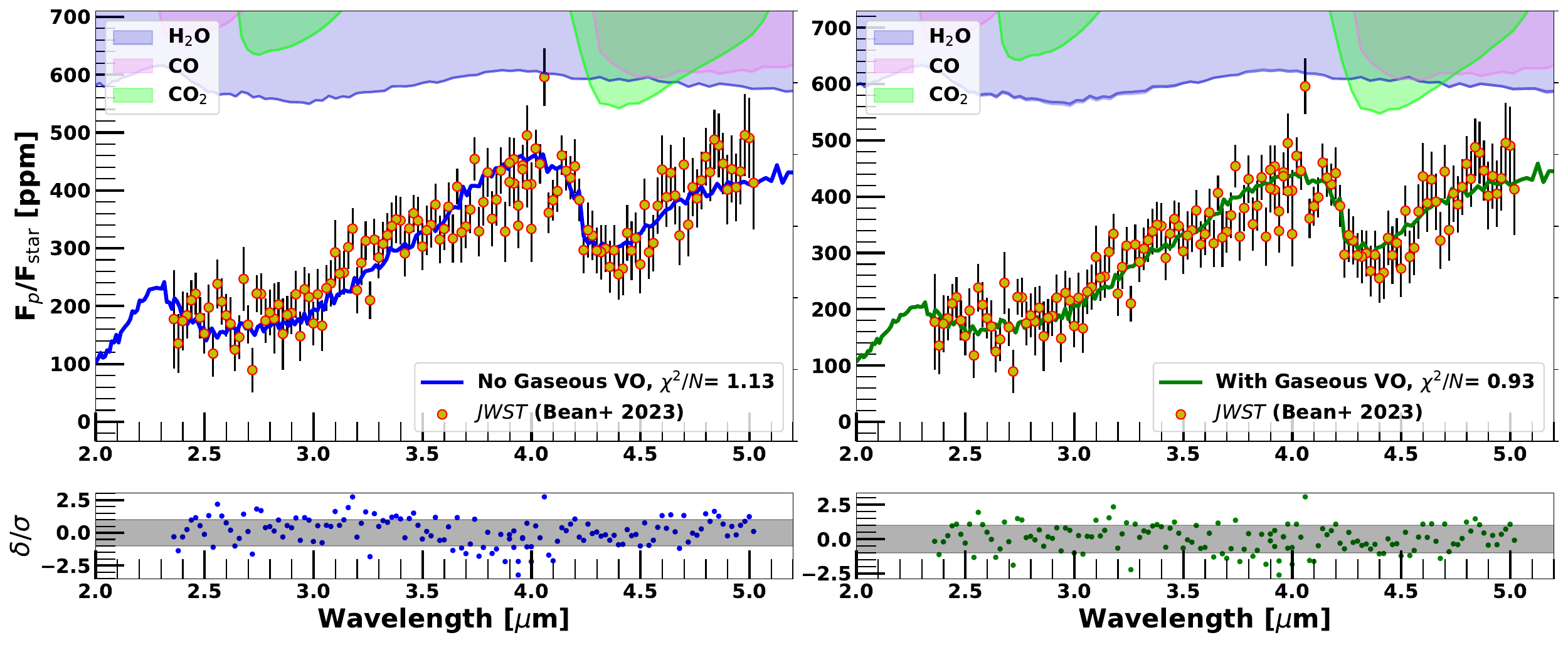}
    \centering
    \caption{The top left panel shows the thermal emission spectrum data from \textit{JWST} with the best-fit model without VO in the atmosphere. The top right panel shows the same data and the best-fit model including VO. Both of the plots in the top row also show the $\tau = 1$ pressure levels for $H_2O$ in blue, $CO$ in pink, and $CO_2$ in green. These are plotted with an inverse y-axis, meaning that the line that dips lowest on the graph is the dominant absorber in the atmosphere. The bottom two panels show the residuals of the best-fit model divided by the noise from the data. The grey shaded region represents values within 1$\sigma$ of the error bars from the \textit{JWST} data.}
    \label{fig:fit_spec}
\end{figure*}

Using the Bayesian grid fitting technique described in \S\ref{sec:gridtrievals}, we fit the NIRCam data of HD 149026b using two separate grids of models -- one without gaseous VO and another with VO. Figure \ref{fig:fit_spec} top left and right panels show the best-fitting spectra obtained from each of these grids compared with the {\it JWST} data. The best-fitting model spectra obtained from the grid without gaseous VO opacity is shown with the blue line in the top left panel in Figure \ref{fig:fit_spec} while the green line in the top right panel shows the best-fitting model spectra obtained from the grid computed with gaseous VO opacity. The bottom left and right panels show the residuals of the model fit divided by the noise in the data from the respective fits shown in the top panels. The gray shaded region in the bottom panels in Figure \ref{fig:fit_spec} mark the regions where the residuals of the fit are within 1$\sigma$ of the error bars in the data. The best-fitting model obtained from the grid without gaseous VO opacity has a reduced $\chi^{2}$  value of about 1.13 whereas the model grid with gaseous VO opacity provides a relatively better fit with a reduced $\chi^{2}$ of 0.93. The retrieved spectra from \citet{bean23} has a reduced $\chi^2$ of 0.94, meaning that our model grid containing VO results in comparable fit. We define the reduced $\chi^{2}$ as $\chi^2/N$, where $N$ is the number of data points.

We also show the $\tau=1$ pressure levels in the left and right top panels in Figure \ref{fig:fit_spec} to identify the dominant gaseous absorbers at each wavelength of the best-fit model spectra. The blue shaded region depicts the $\tau=1$ pressure level for {\water}, the pink for {\co}, and the green for {\cotwo}. The $\tau=1$ pressure levels for each individual gases are shown with an inverted y-axis for illustrative purposes, where the lower part of the y-axis is the upper atmosphere whereas the upper part of the y-axis is the deeper atmosphere. Between 4.1-4.6 $\mu$m, {\cotwo} is the dominant absorber and is responsible for the deep absorption band seen in the data. {\water} remains the dominant absorber between 2.4-4.1 $\mu$m whereas the spectrum is shaped by both {\water} and {\co} between 4.6-5.1 $\mu$m. Our identification of these gases as the dominant absorbers at various wavelength regions remains similar to the results from \citet{bean23}. 

\subsection{Constraints on Atmospheric Parameters}
\begin{figure*}
    \centering
    \includegraphics[width = \textwidth]{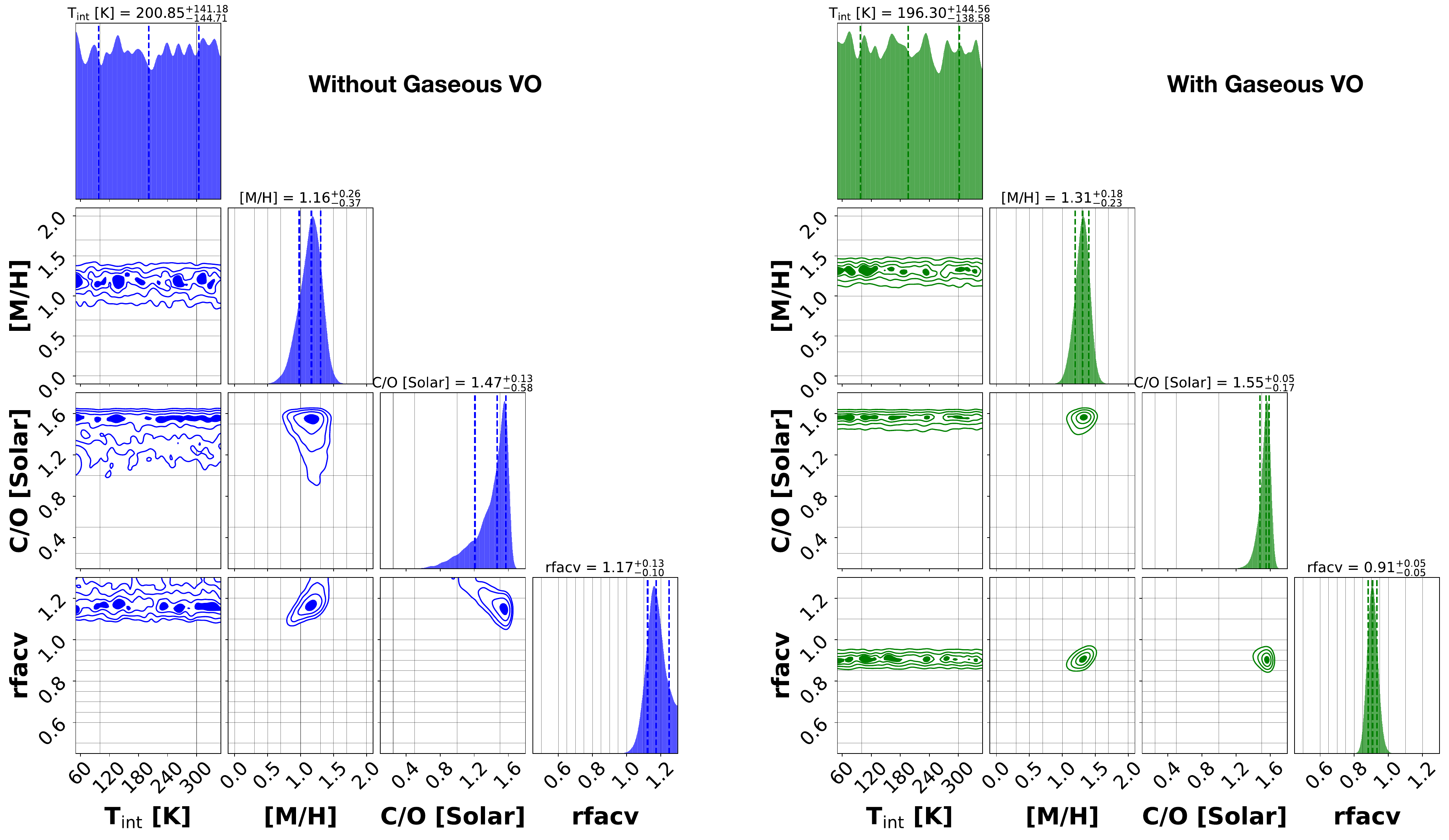}
    \centering
    \caption{Corner plots showing the posterior probability distributions from the Bayesian model fitting for both without (left) and with (right) gaseous VO scenarios are shown here. When gaseous VO is excluded from the atmospheres a median [M/H] of +1.16 is obtained while the median [M/H] increases to +1.31 when gaseous VO is included. The [M/H] is defined relative to solar metallicity. The estimated C/O remains slightly super-solar in both the scenarios and are consistent with each other within 1$\sigma$. The estimated heat redistribution parameter {\rfac} also shows a  change from 1.17 to 0.91 between the no VO and VO case. The {\tint} of the planet remains unconstrained in both scenarios. The gray lines in both the corner plots depict the grid points around which the model grid was computed while the chemistry and the {\tp} profile has been interpolated in between these grid points during fitting the data.}
    \label{fig:corner}
\end{figure*}

Figure \ref{fig:corner} shows the posteriors on the atmospheric parameters -- {\tint}, metallicity, C/O, and \rfac\ obtained with the Bayesian interpolated grid fitting process. The corner plot shown on the left side of Figure \ref{fig:corner} corresponds to the scenario where the atmosphere lacks VO whereas the corner plot in the right has the amount of gaseous VO predicted by thermochemical equilibrium. The gray lines in Figure \ref{fig:corner} show the grid-points for each parameter in our grid. The atmospheric {\tp} profiles and chemistry is interpolated when the Bayesian sampler draws parameters from positions in between these grid points.

It is clear from Figure \ref{fig:corner} that obtaining constraints on {\tint} is not possible with thermochemical equilibrium models. The last panels in Figure \ref{fig:models} makes it clear the {\tint} only affects the atmosphere at pressures deeper than $\sim$ 0.5 bars which cannot be probed with the available observations under the assumptions of thermochemical equilibrium. When the data is fitted with models without any gaseous VO opacity, we obtain a median metallicity of [M/H]= 1.16$^{+0.26}_{-0.37}$. The obtained median metallicity corresponds to a metallicity of 14.4$\times$ solar. When we fit the data with models with gaseous VO, we obtain a slightly higher median metallicity of [M/H]= 1.31$^{+0.18}_{-0.23}$ which corresponds to a metallicity of 20.4$\times$ solar. In both cases, the atmosphere appears to be more metal-rich than both the host star and the sun. The estimated metallicity of the host-star is +0.25, therefore the planet appears to be about 10$\times$ more metal-enriched than the host star.

However, our constraint on the metallicity of HD 149026b is much lower than the constraints obtained by \citet{bean23} from chemically consistent Bayesian retrieval analysis. Based purely on the atmospheric mass-metallicity relationship estimated in the literature \citep[e.g.,][]{wellbanks19,thorngren16,kreidberg14,fortney13}  HD 149026b is predicted to have an atmospheric metallicity between 2-30$\times$ solar metallicity. Our metallicity constraint is consistent with this prediction whereas the metallicity estimated in \citet{bean23} is not. We discuss this further \S\ref{sec:massmet}. We believe the main reason behind the difference in the metallicity estimate is the difference in the behavior of the {\tp} profile obtained from our self-consistent RCTE models and the retrieved {\tp} profile obtained by \citet{bean23}, but additional observations at wavelengths outside those currently available are necessary to determine which of these metallicities produce better fitting spectra. The parameterized {\tp} profile retrieved by \citet{bean23} has an isothermal nature at pressures smaller than 10 mbars (and at pressures greater than 0.1 bars). While the rest of the spectra is primarily sensitive to pressures between 10 mbars and 0.1 bars, the depth of the {\cotwo} absorption band between 4.1--4.6 $\mu$m is quite sensitive to the {\tp} profile at pressures smaller than 10 mbars \footnote{The published online correspondence between \citet{bean23} and the two paper referees makes clear that all were aware of this issue, and a retrieval was performed with a more flexible \tp profile.  A largely isothermal upper atmosphere was still retrieved, with a high CO$_2$ abundance, as such a profile yielded the statistical best fit.} as we will quantify via contribution functions in Figure \ref{fig:contribution}. 

If an isothermal {\tp} profile is retrieved for pressures lower than 10 mbars, the {\cotwo} abundance needs to increase to still account for the same depth of the {\cotwo} feature seen in the data. As the {\cotwo} abundance increases strongly with metallicity, a small increase in the required {\cotwo} abundance due to an isothermal {\tp} profile at low pressures might result in a sharp rise in the required atmospheric metallicity to fit the data. The RCTE models presented in this work, on the other hand, do not show the isothermal {\tp} profile obtained by the retrieval analysis in \citet{bean23}. As a result, a smaller {\cotwo} abundance can create enough difference in the brightness temperature in wavelengths inside and outside the {\cotwo} absorption band. Therefore, our constraints on metallicity are an order of magnitude smaller than the constraints obtained by \citet{bean23}.

The constraints on the C/O ratio for HD 149026b's atmosphere are also depicted in Figure \ref{fig:corner}. In both the without VO and with VO scenarios, subsolar C/O ratios are strongly ruled out by the data. In the fit without gaseous VO, the median C/O ratio obtained is 1.47$\times$solar. This corresponds to C/O= 0.673. 
However, the posterior distribution of C/O from the grid without VO shows a small probability of a solar C/O value as well. A supersolar C/O ratio of 0.687 is strongly favored by the models with VO. Our grid includes another supersolar C/O grid point for models with C/O= 0.916. These models show thermal inversions as shown in Figure \ref{fig:models} and therefore we exclude them from our analysis. The estimated C/O value from the retrieval analysis is 0.84, which is slightly higher than our estimate (\citep{bean23}). However, this difference may well be due to the lack of another grid point in our models nearer to the C/O ratio estimated by \citet{bean23}.

Our analysis also constrains the heat redistribution parameter, \rfac, for HD 149026b. \rfac\ is found to be 1.17$^{+0.13}_{-0.10}$ when the data is fit with models without VO. This reflects a scenario where the stellar heat incident upon the day side of HD 149026b is entirely re-radiated by the day side with very little heat transport even on the day side. This represents a scenario where the winds have a transport timescale slower than the radiative timescale of its atmosphere. In models with VO, a slightly lower and more tightly constrained heat redistribution parameter of $0.91^{+0.05}_{-0.05}$ is obtained. This represents a scenario where a small fraction of the incident energy is transported to the planet's night side. Both of these values are far away from 0.5, which represents the scenario where both the day and the night side of the planet re-radiate half of the stellar energy received. Therefore, our analysis suggests that HD 149026b should have large day-night temperature contrasts. These results are somewhat in line with phase curve measurements of HD 149026b obtained with {\it Spitzer} by \citet{zhang18_hd149206}. Using the toy model of heat recirculation from \citet{cowan11}, \citet{zhang18_hd149206} measured the heat recirculation $\epsilon$ parameter for HD 149026b to be around $\sim$ 0.3. A value of 0 for $\epsilon$ represents no heat recirculation at all and is similar to our value of 1.33 for the \rfac. If $\epsilon$ is 1 then that represents the scenario that the day and the night side of the planet are at the same temperatures and corresponds to a value of 0.5 for our heat redistribution parameter. Measurements from \citet{zhang18_hd149206} show that the heat recirculation in HD 149026b's atmosphere between the day and the night side is rather low similar to our findings.  A spectroscopic phase curve with \emph{JWST} would be an important followup to the present dayside spectrum.

\subsection{Thermal Contribution Function and Atmospheric Chemistry}
\begin{figure*}
    \centering
    \includegraphics[width = \textwidth]{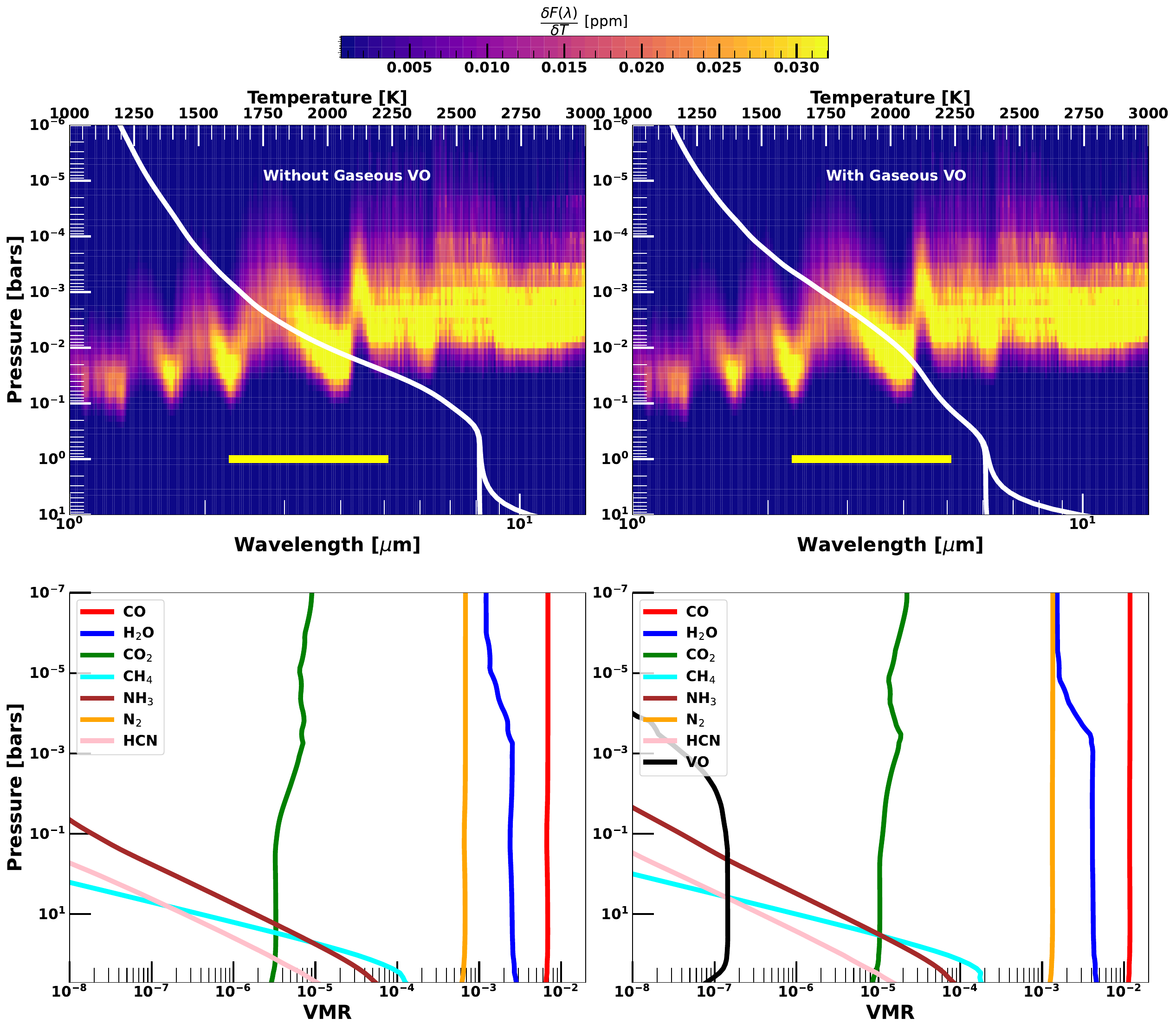}
    \centering
    \caption{The top panels show the temperature-pressure profile for the best-fit models without gaseous VO on the left and with gaseous VO on the right. The contribution of each atmospheric layer to the thermal emission of the planet, calculated using Equation \ref{CF}, is also shown in the two top panel plots. The pressure layers that are more yellow contribute more than the darker regions. The yellow bar in the two top panels depict the wavelength range of the {\it JWST} observations used in this work. The bottom panels show the atmospheric abundances of crucial atmospheric gases in the best-fit models assuming {\tint}=100 K without VO on the left and with VO on the right.}
    \label{fig:contribution}
\end{figure*}

Thermal contribution functions are useful tools to understand which levels of the atmosphere are responsible for emission at a particular wavelength. We quantify the contribution of a particular atmospheric layer to the emitted flux at a certain wavelength using,
\begin{equation}\label{CF}
    CF(\lambda,P) = \dfrac{\partial{F^{\rm out}}(\lambda)}{\partial{T(P)}}
\end{equation}
where $CF(\lambda,P)$ is the contribution function of the layer with pressure P at wavelength $\lambda$, $F^{\rm out}$ is the emitted flux at the top of the atmosphere at wavelength $\lambda$, and $T(P)$ is the temperature of the atmospheric layer with pressure $P$. This equation mainly captures the change in the emitted flux of the planet caused by a tiny perturbation to its temperature. If a certain layer is contributing a large fraction of the emitted flux by the planet, then a tiny change in its temperature should cause a large change in the outgoing flux and that would result in a large value for $CF(\lambda,P)$. On the other hand, if a layer is contributing negligible flux to the emitted flux then a perturbation of its temperature shouldn't cause the outgoing flux to change much. 

To compute the contribution function from our best-fit models, we perturb the temperature of each model atmospheric layer by an amount $\delta{T}$=10$^{-5}T$ where $T$ is the unperturbed temperature of the atmospheric layer. We compute the emitted flux with this perturbed layer temperature at each wavelength and compute its difference from the unperturbed outgoing flux. This allows us to estimate $CF(\lambda,P)$ directly without any functional forms. Figure \ref{fig:contribution} top left and right panels show the heat maps depicting the contribution function obtained from our best-fit models without and with VO, respectively. The best-fit {\tp} profiles obtained in each case are also overplotted on each of the two top panels in Figure \ref{fig:contribution}. 

It is clear that between 2-4.1 $\mu$m, most of the observed flux is being emitted by the atmosphere between 1 mbar and 0.1 bar in both scenarios. The best-fit atmospheric {\tp} profiles have a strong slope from $\sim$ 1500 to 2000 K in this pressure range for both models. Interestingly, the retrieved parameterized {\tp} profile obtained by \citet{bean23} also has a similar slope between very similar temperatures in the same pressure range, as we will see in Figure \ref{fig:jwst_spitzer}. In wavelengths within the {\cotwo} absorption band between 4.1-4.6 $\mu$m, the contribution function moves to pressures between 10 mbar to 0.1 mbar where the RCTE best-fit models are slightly colder than 1500 K but are still within the uncertainty envelopes of the retrieved profiles in \citet{bean23}. The {\tp} profiles of our best-fit models differ significantly from the retrieved profiles from {\tp} above this pressure range. However, it is clear from Figure \ref{fig:contribution} that this {\it JWST} spectrum is not enough to resolve the uncertainty about the nature of the {\tp} profile a pressures below 0.1 mbar. However, the longer wavelengths have relatively higher contribution from pressures smaller than 0.1 mbar than this wavelength range.

The bottom left and right panels in Figure \ref{fig:contribution} show the abundance profiles of important atmospheric gases in the best-fit models obtained in our analysis without and with VO, respectively. As the best-fit metallicity obtained by each of set of these models are slightly different, there are large differences in the abundances of {\cotwo} and {\co} between the best-fit models with and without VO. Also, the best-fit model with VO has a VO volume mixing ratio of $\sim$ 10$^{-7}$ at pressures greater than $\sim$ 1 mbar but the VO abundance falls below this pressure due to V- condensation. The inferred atmospheric abundances are also very different from the abundances inferred by \citet{bean23} due to the large difference in constrained atmospheric metallicity between the two analyses.

\subsection{Comparison with Other Measurements}\label{sec:PLATON}
\begin{figure*}
    \centering
    \includegraphics[width = \textwidth]{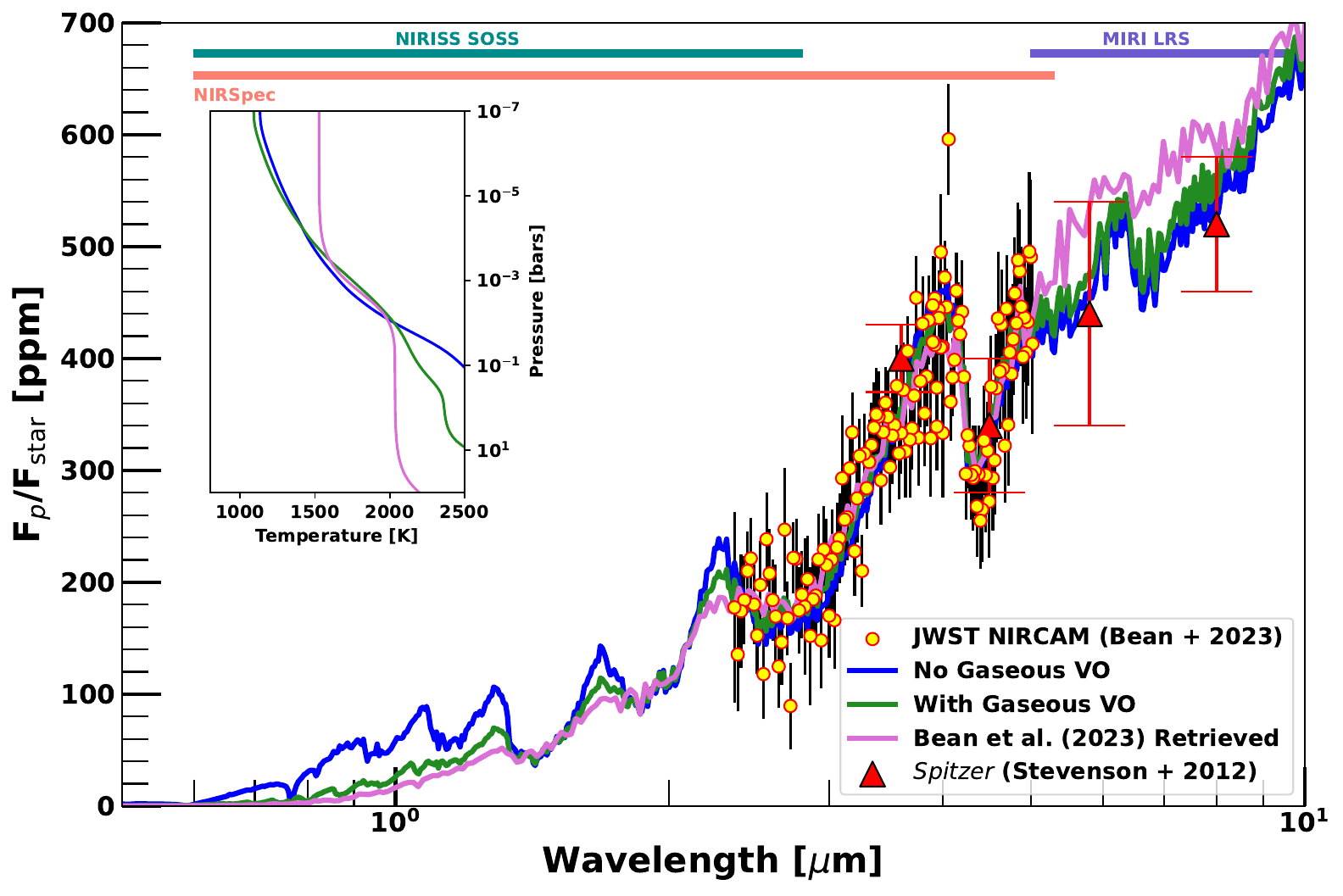}
    \centering
    \caption{Thermal emission spectra for our best-fit models compared with the \citet{bean23} retrieved best-fit spectrum, the \textit{JWST} NIRCam data, and the \textit{Spitzer} data from \citet{stevenson12}. Our model without VO is shown in blue, with VO is shown in green, the \citet{bean23} retrieved spectrum is shown in pink, the \textit{JWST} data is shown with yellow points, and the \textit{Spitzer} data is shown with red triangles. The inset shows the median retrieved \tp profile from \citet{bean23} in pink compared with the with and without VO best fit RCE model \tp profiles from this work. The blue profile is without VO whereas the green model includes VO. The wavelength coverage of various {\it JWST} instrument modes are also shown at the top.}
    \label{fig:jwst_spitzer}
\end{figure*}
With our Bayesian grid fitting technique, we only fit the {\it JWST} NIRCAM data presented in \citet{bean23}. However, we also compare our models and the retrieved spectra from \citet{bean23} with measurements of the planet's thermal emission at longer wavelengths obtained with {\it Spitzer} by \citet{stevenson12}. To reproduce the median spectra retrieved by \citet{bean23}, we use the PLATON tool \citep{zhang2020} along with the PLATON opacities and the atmospheric parameters retrieved by \citet{bean23} instead of \texttt{PICASO} to avoid any systematic differences in opacities used between the two models. Figure \ref{fig:jwst_spitzer} shows the comparison between three best-fit models with the {\it JWST} data in yellow and the {\it Spitzer} data shown with red triangles. The blue and the green models show the best-fit RCTE models obtained with our analysis without and with VO, respectively. It is clear that the main difference between these two models appear at wavelengths shorter than 2.3 $\mu$m. The lower metallicity of the best-fit RCTE model without VO, causes it to have higher fluxes in between the {\water} absorption bands in this wavelength region due to smaller amount of {\water} in this best-fit model than the best-fit model with gaseous VO. The differences between these two best-fit models become much larger at wavelengths smaller than 1.2 $\mu$m due to the additional VO opacity in the models with gaseous VO shown with the green line. The absence of VO causes much higher flux in the without-VO model in these short wavelengths compared to the with-VO model as VO has very high optical opacities. 

Both of these RCTE models are consistent with all the {\it Spitzer} observations in and outside the wavelengths covered by the NIRCam spectra. As also found by \citet{zhang18_hd149206}, super-solar metallicity models are consistent with these {\it Spitzer} measurements. Figure \ref{fig:jwst_spitzer} also shows that the retrieved spectra from \citet{bean23} are consistent with the {\it Spitzer} measurements inside the wavelength covered by the NIRCAM data but lie at the boundary of consistency or inconsistency with the 1$\sigma$ uncertainties of the {\it Spitzer} measurements at longer wavelengths.

Figure \ref{fig:jwst_spitzer} and \ref{fig:contribution} makes it clear that more observations at shorter wavelengths or longer wavelengths than the NIRCAM wavelength coverage are needed to differentiate between these two scenarios of the non-isothermal upper atmosphere with lower metallicity predicted by RCTE models and isothermal upper atmosphere with much higher metallicity predicted from Bayesian retrievals. Observations of transmission or emission spectra at wavelengths near or shorter than 1 $\mu$m will also be helpful in determining whether this planet has gaseous VO in its upper atmosphere or not.

\section{Discussion}\label{sec:discussion}

\subsection{Self-consistent Forward Models and Bayesian Retrievals}

\begin{figure}
    \centering
    \includegraphics[width = 0.5\textwidth]{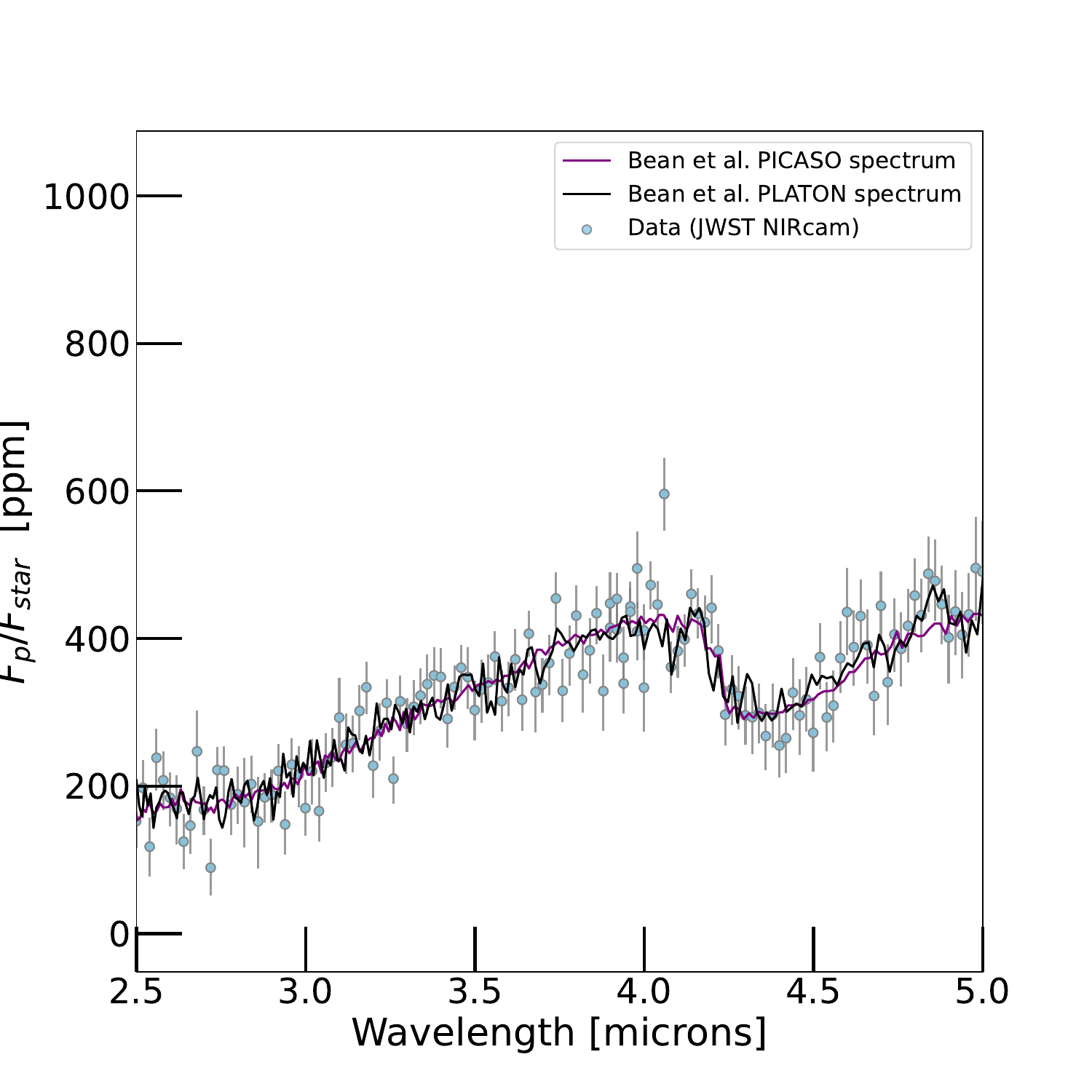}
    \centering
    \caption{Thermal emission spectrum from PICASO using the \citet{bean23} atmospheric parameters and abundances in purple along with the spectrum from PLATON using the \citet{bean23} atmospheric parameters in black. The data from \textit{JWST} NIRCam is also included in blue. Both the PICASO and PLATON spectra are rebinned to a spectral resolution of 300. At this resolution, the median of the difference between the two spectra is 17.7 ppm. This difference is potentially dominated by three key factors: 1) PLATON spectra was created with the default resampled opacities at R $\sim$ 1000 whereas the \texttt{PICASO} spectra was calculated using resampled opacities at R $\sim$ 10000, 2) the sources of several key line lists are different between PLATON and \texttt{PICASO}, and 3) the opacities in PLATON are interpolated from a grid of 390 P-T points whereas the opacities in \texttt{PICASO} are interpolated from a denser grid of 1460 P-T points. We note that PLATON does provide options for higher resolution resampled opacities.   }
    \label{fig:specbeans}
    \end{figure}

Atmospheric retrievals provide an important data-driven pathway to constrain fundamental atmospheric properties of planets. However, this work shows that it is also necessary to compare the retrieved constraints with self-consistent forward models in order to put the retrieval results in the context of our theoretical understanding of exoplanetary atmospheres. In this case, the retrievals done by \citet{bean23} on the \textit{JWST NIRCAM} data estimate an atmospheric metallicity of $59-276 \times$ solar whereas our self-consistent modeling analysis points to a much lower metallicity estimate between 10-20$\times$solar metallicity for HD 149026b. We suggest that the root cause of this is the difference in the retrieved upper atmosphere \tp profile and those predicted from 1D self-consistent models. It is difficult to differentiate between these two scenarios with the currently available data for HD 149026b, but this also shows why it is important to interpret observational data both with retrievals and self-consistent forward models. As a check on potential systematic differences between \texttt{PLATON} and \texttt{PICASO}, we used the medianed retrieved atmospheric parameters and molecular abundances from \citet{bean23} to create an emission spectrum with PICASO to compare with the PLATON spectrum discussed in Section \ref{sec:PLATON}. These can be seen in Figure \ref{fig:specbeans}. This resulted in spectra with a $\chi^2$ value of  0.98 and 0.94 for the PICASO and PLATON spectra respectively. The differences between our atmospheric parameters and those found by \citet{bean23}, specifically at wavelengths outside of those covered by the \emph{JWST} spectra as shown in Figure \ref{fig:jwst_spitzer}, support the need for additional observations of HD 149026b in order to validate which of these models is more reasonable.

While retrievals can be used to provide statistical constraints on various atmospheric parameters, self-consistent models can be used to interpret these constraints and investigate what the constraints indicate about the nature of the planet and its atmosphere. It is now possible to produce very large planet-specific self-consistent model grids covering large parts of relevant parameter space. Such large grids can also be used to fit observational data and the results can be compared with retrieved results, as has been done in this work. The similarities or differences in the results can then be used to study the missing complexity in each approach and to suggest  future observations of planets to break degeneracies/discrepancies between the two methods. A synergy between the retrieval and forward modeling approaches is also perhaps the best way to inform the assumptions within the forward models from observational data. For example, if the upper atmosphere of HD 149026b is indeed found to be as isothermal and hot with future observations as found by the retrieval study, then it is critical to understand the missing physics/complexity in the self-consistent models be identified and implemented. On the other hand, if the retrieval results are negated with future observations, then we would need to reassess the subtleties involved in retrieval studies like parametrization of \tp profiles and their effect on retrieved results. This would be an excellent use of pioneering {\it JWST} observations to inform our physical understanding of exoplanetary atmospheric physics and to improve our interpretation techniques.

Any 1D model of an atmosphere, be it RCTE or a parameterized retrieval profile, is a simplification of a 3D reality.  \citet{seager05} pointed out that lower pressure regions may be closer to pure radiative equilibrium, making 1D models more appropriate, while high pressure regions may be dominated by advective transport.  Future word on GCMs for the planet, building off the work of \citet{zhang18_hd149206}, which showed large day-night temperature contrasts for their $30 \times$ solar models, but not at $1\times$, is certainly warranted given these new \emph{JWST} spectra. 

\subsection{Night Side}

The day-side effective temperature of the best-fit 1D forward model with gaseous VO was found to be 1952 K. This temperature is consistent within 1$\sigma$ with the day-side brightness temperature derived by \citet{zhang18_hd149206} from the {\it Spitzer} 3.6 $\mu$m channel phase curve measurements. A comparison of the day-side emitted flux calculated using this effective temperature with the total incident flux on the planet implies that $\sim$88\% of the incident energy on HD 149026b is being re-emitted from the day-side of the planet and $\sim$ 1\% is being reflected back to space. The the rest ($\sim$11\%) is being circulated to and re-emitted from the night side. If $\sim$11\% of the total incident energy is being recirculated to the night-side, then the night-side of the planet should have an effective temperature of 1159 K. This corresponds to a day-to-night temperature contrast of about 793 K. Both the night-side effective temperature and the day-to-night side temperature contrast implied from our work are consistent within 1$\sigma$ with the {\it Spitzer} 3.6 $\mu$m phase curve measurements reported in \citet{zhang18_hd149206}. Both the day-side and night-side effective temperature derived in this work are very slightly inconsistent with the 1$\sigma$ limits on the wavelength-averaged day and night brightness temperature of HD 149026b derived in \citet{zhang18_hd149206}.

Our best-fit model without VO has an \rfac value of 1.17 and a day-side effective temperature of 2115 K. This scenario, with \rfac\ $>1$, implies no heat transport to the night-side of the planet and therefore seems less plausible physically. We take this as circumstantial evidence that VO is present in the atmosphere.  We suggest high-resolution spectroscopic search for the molecule and addition \emph{JWST} observations to constrain the \tp\ profile. We note that the equilibrium temperature of HD 149026b is 1694 K, which is very close to the 1730 K equilibrium temperature around which an abrupt rise in day side brightness temperature has been recently found by \citet{deming23}.


\subsection{Mass-Metallicity Relation}\label{sec:massmet}

Gas giant planets in the solar system follow an inverse mass- atmospheric metallicity relationship where lower mass planets are found to be more metal-rich relative to their host stars than massive planets. Such a trend has also been tentatively found for exoplanets \citep[e.g.,][]{wellbanks19,thorngren16,kreidberg14,ers_g395h,ers_prism,ers_nircam,ers_niriss} and suggested from formation theory \citep{fortney13}. However, \citet{bean23} results found HD 149026b's metallicity to be quite inconsistent with this mass-metallicity trend, where it was much more metal-enriched than expected for its mass. The metallicity estimate from the self-consistent modeling approach in this work, on the other hand, makes HD 149026b consistent with the broader mass-atmospheric metallicity relationship. This consistency cannot be used to claim that one result is favored over the other but instead points to the necessity of future {\it JWST} observations covering a broader wavelength range to break the degeneracy between the two approaches used to interpret the data presented in \citet{bean23}.




\section{Conclusions}\label{sec:conclusion}
By creating a grid of 1D RCTE atmosphere models for HD 149026b, we were able to infer its atmospheric properties through comparison with \textit{JWST NIRCAM} eclipse spectroscopy. As HD 149026b belongs to a part of the parameter space where gaseous TiO and VO may or may not be condensed, we have considered both the case of an atmosphere that includes gaseous VO and one that does not. We have also found that the inclusion of gaseous TiO causes large temperature inversions, which is not compatible with the {\it JWST} observations.  Both with and without  VO cases can fit the {\it JWST} and {\it Spitzer} observations well. We draw the following conclusions in this work.
\begin{enumerate}
    \item \textbf{Atmospheric Metallicity:} Our models resulted in an estimated atmospheric metallicity of $14^{+12}_{-8}\times$ solar when we assume the atmosphere contains no gaseous VO and $20^{+11}_{-8}$ $\times$ solar when we assume the atmosphere contains gaseous VO. The constrained metallicities from each scenario are consistent with each other within 1$\sigma$. This is significantly lower than the retrieved metallicity from \citet{bean23} of $59 - 276$ $\times$ solar.  We attribute this difference to the difference between the slope of the retrieved {\tp} profile and the {\tp} profiles calculated with the self-consistent models. The planet is $\sim10\times$ more metal-enriched than its host star, which has a metallicity of 1.8$\times$ solar metallicity.
    
    \item \textbf{Atmospheric Recirculation:} We find that the atmosphere has a high heat redistribution factor, suggesting low atmospheric recirculation. Our heat redistribution factors, \rfac, of $0.91^{+0.05}_{-0.05}$ and $1.17^{+0.13}_{-0.10}$ for atmospheres with and without VO, respectively, indicate that the majority of the energy from the star is retained on the day side of the planet, with little redistribution of heat to the night side.  The potentially implausibility of \rfac\ $>1$ for the no-VO models suggests that this molecule, which condenses at temperatures 200 K lower than TiO, may be found in the planet's atmosphere.

    \item \textbf{Gaseous TiO and VO:} The inclusion of gaseous VO in the atmosphere moderately impacts the best-fit atmospheric metallicity and also yields a more plausible heat recirculation factor.  We suggest further observations to search for the molecule either directly (at high spectral resolution) or indirectly (via a better constrained \tp\ profile).
    
    \item \textbf{C/O Ratio:} We estimate the C/O ratio to be roughly 0.67, or perhaps modestly higher, for atmospheres both with and without gaseous VO.  Within the confines of our chemistry grid, C/O ratios $<0.9$ and $>0.22$ are preferred, but we cannot tightly constrain this ratio. The upper limit of the constraint on C/O comes from the lack of a temperature inversion (Figure \ref{fig:models}) 
    
    \item \textbf{Mass-Metallicity Relationship:} The atmospheric metallicity for HD 149026b constrained in this work is consistent with the metallicity predicted for the planet from the atmospheric mass-metallicity relationship estimated in the literature. 

\end{enumerate}
Our inferred atmospheric parameters like atmospheric metallicity and {\tp} profiles are significantly different from the retrieval results from \citet{bean23}. As it stands with the data we have now, we are unable to definitively say which is more a reasonable view of the atmosphere. We have shown that future observation in the 0.8-2 $\mu$m by either \textit{JWST} or the \textit{Hubble Space Telescope} is necessary to determine whether the atmosphere contains gaseous VO or not. Observations in longer infrared wavelengths with instruments like {\it JWST MIRI} will also be crucial to determine which scenario more accurately describes the atmosphere of HD 149026b -- the inferred atmospheric structure and composition in this work, or the retrieved {\tp} profile and chemistry presented in \citet{bean23}, as MIRI probes relatively low pressures. Both of these observations are crucial, as they can lead us to find missing physical/chemical processes in our 1D atmospheric models (e.g., additional opacity) and can also highlight the improvements needed in retrievals studies of exoplanet eclipse spectroscopy data.

\section{Acknowledgments}
AG thanks the UC Regent's Fellowship as well as the Cal Bridge Astro Fellowship for supporting her in this work. SM thanks the UC Regents Fellowship award for supporting him for this work. JJF acknowledges the support of NASA XRP grant 80NSSC19K0446 and 
\emph{JWST} grant JWST-AR-01977.003-A. We acknowledge use of the \emph{lux} supercomputer at UC Santa Cruz, funded by NSF MRI grant AST 1828315. We thank the anonymous referee for their feedback on our manuscript.
We also thank Linsey Wiser for valuable model comparisons between \texttt{PICASO} and ScChimera.
 
{\it Software:} \texttt{PICASO 3.0} \citep{Mukherjee22}, \texttt{PICASO} \citep{batalha19}, pandas \citep{mckinney2010data}, NumPy \citep{walt2011numpy}, IPython \citep{perez2007ipython}, Jupyter \citep{kluyver2016jupyter}, matplotlib \citep{Hunter:2007}, the model grid has been formally released via Zenodo and can be accessed at \url{doi:10.5281/zenodo.11062209}.

{\it Data:} The \textit{JWST} datasets used in this paper are from the Mikulski Archive for Space Telescopes (MAST) at the Space Telescope Science Institute (\textit{StScI}). The uncalibrated data can be accessed at \url{doi:10.17909/qn84-n051}.

\bibliography{arxiv}{}
\bibliographystyle{apj}



\end{document}